\documentclass[prl,floatfix,twocolumn,showpacs]{revtex4}
\usepackage{amsmath}
\usepackage{amsfonts}
\usepackage{mathrsfs}
\usepackage{epsfig}

\usepackage{graphicx}
\usepackage{dcolumn}
\usepackage{amsmath}

\def\be{\begin{equation}}
\def\ee{\end{equation}}

\begin{document}
\title{Modular organization enhances the robustness of attractor
network dynamics}
\author{Neeraj Pradhan$^{1,2}$, Subinay Dasgupta$^3$ and  Sitabhra
Sinha$^1$}
\affiliation{%
$^1$The Institute of Mathematical Sciences, CIT Campus, Taramani,
Chennai 600113, India.\\
$^2$Department of Physics, Birla Institute of Technology \& Science,
Pilani 333031, India.\\
$^3$Department of Physics, University of Calcutta, 92 Acharya Prafulla
Chandra Road, Kolkata 700009, India.}
\begin{abstract}
Modular organization characterizes many complex networks occurring in
nature, including the brain. In this paper we show that modular
structure may be responsible for increasing the robustness of certain
dynamical states of such systems. In a neural network model with
threshold-activated
binary elements, we observe that the basins of attractors,
corresponding to patterns
that have been embedded using a learning rule,
occupy maximum volume in phase space at an optimal 
modularity. Simultaneously, the convergence time
to these attractors decreases as a result of cooperative dynamics
between the modules. The role of modularity in increasing global
stability of 
certain desirable attractors of a system may provide a clue to its
evolution and ubiquity in natural systems.
\end{abstract}

\maketitle

\newpage
An ubiquitous property of complex systems is their modular
organization~\cite{Hartwell99}, characterized by communities
of densely connected elements with sparser connections
between the different communities~\cite{Girvan02}.
In the biological world, modules
are seen to occur across many length scales, from the intra-cellular
networks of protein-protein interactions~\cite{Schwikowski00,Rives03} 
and signaling pathways~\cite{Lauffenburger00} to
food webs comprising multiple species populations~\cite{Krause03}.
Although such groupings are primarily defined in terms of the structural
features of the network topology, in several instances distinct
modules have also been associated with specific functions. 
Indeed, in the case of the brain, modular organization at the
anatomical level has long been thought to be paralleled at the
functional level of cognition~\cite{Fodor83}.
By observing the effects of isolating or disconnecting different brain  
areas on the behavior of subjects, the functional specialization of
spatially distinct modules have been established at different length
scales~\cite{Gazzaniga89} -
from hemispheric specialization to minicolumns comprising a few
hundred cells which have been proposed as the basic information
processing units of the cerebral
cortex~\cite{Mountcastle75,Buxhoeveden02}.
More recently, the analysis of neurobiological data using 
graph theoretic techniques~\cite{Bullmore09} has further established the modular
nature of inter-connections between 
different areas of the mammalian cortex. The structural modules
revealed by tracing the anatomical
connections in mammalian brains~\cite{Hilgetag00,Chen08} are complemented by
the observation of functionally defined networks having modular
character~\cite{Achard06,Meunier09}.
Such functional networks have been reconstructed
from MRI and fMRI experiments on both
human~\cite{Salvador05} and non-human~\cite{Schwarz08}
subjects, by considering two brain areas to be connected if they are
simultaneously active when the subject performs a specific behavioral task. 

%
The wide-spread occurrence of modularity prompts the question as to
why this structural organization is so ubiquitous~\cite{Pan07}. One
possible reason is that it enhances communication efficiency by
decreasing the average network path length while allowing high
clustering to help localize signals within subnetworks~\cite{Pan09}.
However, of more interest is the possibility that modularity may play
a crucial role in the principal function of the system, viz.,
information processing in the case of brain networks. This possibility 
has been
investigated in detail for the somatic nervous system of the nematode
{\em C. elegans}~\cite{Pan10}. It is therefore intriguing to speculate
whether modularity is responsible for efficient information processing
in brains of more evolved organisms, the mammalian
cortex in particular. To explore this idea further we can study the effect of modular
structure on the dynamics of
attractor network models with threshold-activated nodes, which exhibit
multiple stable states or ``memories"~\cite{Hopfield82,Amit89}. These
models were
originally developed to understand how the nervous system communicates
among its component parts and learns associations between different
stimuli so that a memorized pattern can be retrieved in its entirety from a
small part or a noise-corrupted version of it given as input
(``associative memory''). 
Indeed, recent
experiments indicate that the spatiotemporal activation dynamics in
neocortical networks converge to one of several different 
persistent, stable patterns which resemble the behavior observed in such
models~\cite{Cossart03}. However, the properties of attractor networks
are of more general interest and have been used to understand
systems outside the domain of 
neurobiology, as for example, the network involved
in intracellular signaling where communication between molecules
within a cell take place through multiple interacting
pathways~\cite{Bray95,Bray03}.
In the attractor networks, desired patterns are stored by
using a {\em learning rule} to determine the connection weights
between the nodes. This ensures that the update (or recall) 
dynamics of the network makes it
converge to these pre-specified dynamical states when an input initial
state of the system is transformed into an output state defined over
the same set of nodes by the collective dynamics of the network.
Using such simplified models have the
advantage of making the observed phenomena simpler to analyze and also
to obtain results that are independent of specific biological details of
different types of neurons and synaptic connections.
%
%
%

In this paper we show that if we want to store $p$ (say)
patterns in a network with a given number of nodes and
links, then the
convergence to an attractor corresponding to any of the stored
patterns (i.e., recall)
will be most efficient when the network has an optimal modular
structure, provided the number of patterns is not too large ($p <
p_{max}$).
If the degree of modularity is increased or decreased from the optimum
value,
the reliability with which the patterns are recalled decreases. This
optimal
efficiency of recall originates from the network dynamics itself. Some
of the
modules converge quickly to attractors corresponding to parts of
stored patterns and then help
other modules to reach the attractor corresponding to the entire
stored pattern through interactions via intermodular links. 
If the modularity is increased
(i.e., if the number of intra-modular links is increased while
reducing the number of inter-modular links to keep the average degree fixed),
the modules cannot interact with each other strongly enough due to
fewer number of intermodular links and the performance of the network
is less efficient. On the other hand, if the modularity is decreased,
the modules themselves become sparsely connected and cannot reach an
attractor rapidly. Also, if we try to store a
larger number of patterns ($p \ge p_{max}$), the advantage of modularity
disappears because of the generation of a large number of spin-glass
states which correspond to spurious patterns.

\begin{figure}
\begin{center}
\includegraphics[width=0.8\linewidth]{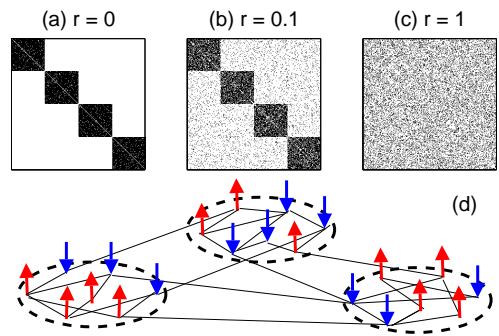}
  \end{center}
    \caption{(a-c) Adjacency matrices ${\bf A}$ defining the network connections 
    at different values of the
    modularity parameter $r$ for $N = 256$ nodes arranged into $n_m =
    4$ modules (average degree $\langle k \rangle = 60$). 
    Starting from a system of isolated clusters (a, $r = 0$), by
    increasing $r$ we obtain modular networks (b, $r = 0.1$)
    eventually arriving at a
    homogeneous network (c, $r = 1$).
    The connection structure of modular networks in the intermediate 
    range $0<r<1$
    is shown schematically in (d). The connection weights have
    different magnitudes and signs.
    }
    \label{fig:schematic}
\end{figure}

The attractor network model we have used to investigate the role of
modularity is constructed such that the $N$ nodes 
comprising it are divided
into $n_m$
modules, each having $n$ ($= N/n_m$) nodes~\cite{Pan09}.
The connection probability between a pair of nodes
belonging to the same module is $\rho_i$, while that between nodes
belonging to different modules is $\rho_o$. 
The modular nature of the network can be
varied continuously by altering the ratio of inter- to intra-modular
connectivity, $r = \frac{\rho_o}{\rho_i} \in [0,1]$, keeping the
average
degree $\langle k \rangle$ fixed (Fig.~\ref{fig:schematic}).  
For $r = 0$, the network
is fragmented into $n_m$ isolated clusters, whereas at $r = 1$,
it is a homogeneous or Erdos-Renyi random network. 
We ensure that the resulting adjacency 
matrix ${\bf A}$ (i.e., $A_{ij} = 1$ if $i, j$ are connected, and 0, 
otherwise) is symmetric. We have explicitly verified that the results
reported below do not change appreciably if ${\bf A}$ is
non-symmetric (corresponding to a {\em directed} network).

The time-evolution of the system is governed by the dynamics of the
variables associated with each node of the network.
An Ising spin $\sigma_i = \pm 1$ is placed at each node which may
represent any binary state variable, such as a two-state neuron (firing=1, 
inactive=$-1$). 
The state of the spins are evaluated at
discrete time-steps using random sequential updating according to the following
deterministic (or zero temperature) dynamics:
\begin{equation}
\sigma_i (t+1) = {\rm sign} \left ({\large \Sigma}_j A_{ij} W_{ij} \sigma_j (t)
\right ),
\end{equation}
where, $W_{ij}$ is the connection strength between neurons $i$ and
$j$. The function sign($z$) = 1, if $z>0$, $=-1$, if $z<0$ and
randomly chosen to be $\pm 1$ if $z=0$. 
The weight associated with each link is evaluated using the Hebbian learning 
rule~\cite{Amit89} for storing $p$ random patterns in an associative network:
\begin{equation}
W_{ij} = \frac{1}{\langle k \rangle} {\large \Sigma}_{\mu} \xi^{\mu}_i
\xi^{\mu}_j, ~W_{ii} = 0,
\end{equation}
$\xi^{\mu}_i$ being the $i$-th component of the $\mu$-th pattern
vector ($\mu = 1, \ldots p $). 
Each of the stored patterns are generated randomly by choosing 
each component to be $+1$ or $-1$ with equal probability. 
Starting from an arbitrary initial state, the network eventually
converges to a time-invariant stable state or attractor.
The overlap of an attractor of the network
dynamics $S^* = \{ \sigma^*_i \}$
with any of the stored patterns can be measured as $m_{\mu} = 
\frac{1}{N}\Sigma_i \sigma^*_i \xi^{\mu}_i$. As we are interested in
the set of all the attractors of stored patterns rather than one
specific pattern, we focus our attention on the maximum overlap with
the stored patterns, $m = {\rm max}_{\mu} |m_{\mu}|$.
To examine the {\em global} stability of the attractors corresponding to the
stored patterns, we use random strings as the initial state of the network 
which should have almost no overlap with any of the
stored patterns, on average. 
The probability $v_g \equiv \langle {\rm Prob}(m >m_o) \rangle$
that such a random initial state eventually almost converges to one 
of the stored patterns, gives an estimate of the overall volume that
the basins of attraction of stored patterns occupy in the
$N$-dimensional network configuration space $\{S\}$. Here $m_o$ is a
threshold for the overlap of the asymptotic stable state
above which the network can be considered to have
recalled a pattern successfully and $\langle \ldots \rangle$ indicates
averaging over many different network configurations {\bf A}, as well
as, pattern ensembles $\{ \xi \}$ and initial states. 
The value of the threshold $m_o$ has been taken
to be $0.95$ for most of the analysis presented here; we have verified
that varying it over a small range does not alter our results.
In a similar way, we can define overlap for each module, $m_\mu
(\alpha) = \frac{1}{n} \Sigma_i \sigma^*_i (\alpha) \xi_i^\mu
(\alpha)$ where the sum is over all spins in the $\alpha$-th module
with
$\alpha=1, \ldots, n_m$ being an index running over the different
modules. The relative size of the basins of attraction at the modular
scale is characterized by the
quantity $v_m = \langle \langle {\rm Prob} (m (\alpha) > m_o)
\rangle_{\alpha} \rangle$, where $m (\alpha) = {\rm max}_{\mu}
|m_{\mu} (\alpha)|$ and $\langle \ldots \rangle_{\alpha}$ indicates averaging
over all the modules.

\begin{figure}
\begin{center}
\includegraphics[width=0.99\linewidth]{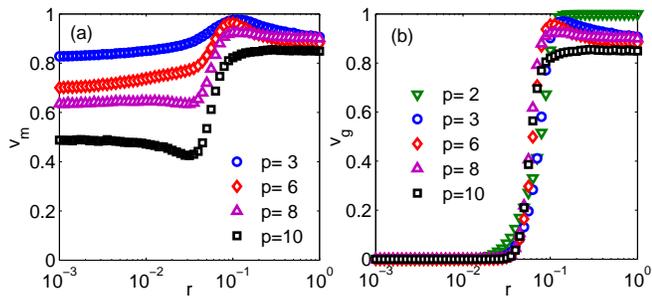}
\end{center}
\caption{Fractional volume of phase space occupied by the basins of
  attraction of the stored patterns in a single module ($v_{m}$) and the
  entire random modular network ($v_{g}$). Note that, when the number
  of stored patterns is within a critical range ($p_{min}=2 < p <
  p_{max}=9$), these quantities show
  a non-monotonic variation with $r$, having a peak around $r_c \simeq
  1/(n_m-1) \sim 0.14$.
  Results are shown for $N=1024$, $n_m = 8$
  and  $\langle k \rangle$ = 120. Different numbers of stored patterns
  $p$ are indicated using various symbols.}
  \label{fig:random_modular_basins}
\end{figure}
We first look at how the total volume of the configuration space
occupied by the basins of attraction for stored patterns $\xi^{\mu}$
changes as the modular character of the network is altered by varying
$r$ for a fixed $\langle k \rangle$. 
Fig.~\ref{fig:random_modular_basins} shows the combined fractional volume of
the phase space
occupied by the basins of
attraction of the stored patterns for the entire network ($v_g$) as
well as for the corresponding sub-patterns in a single module ($v_m$).
Different curves indicate various number of stored patterns $p$.
We immediately notice that while $v_m$ has finite values over the
entire range of $r$, $v_g$ is zero at low values of $r$
where a module is connected to the rest of the network by very few
links, if at all.
The value of $r$ at which $v_g$ starts rising from 0 
appears to be
independent of the number of stored patterns $p$.
Below this value of $r$, the connectivity between the
modules is insufficient to recall the entire stored pattern, even
though individual modules may have complete overlap with different 
stored patterns. 
To explain the situation, we can decompose each stored pattern
in terms of $n_m$ sub-patterns defined over the different modules, viz., 
$\xi^{\mu} = \{ \xi^{\mu} (\alpha)\}$, where $\alpha = 1, \ldots, n_m$.
Starting from a random initial state, a module $\alpha$ may converge to an
attractor corresponding to any of the $n_m$ different subpatterns 
$\xi^{\mu} (\alpha)$. As the recall dynamics within each module is 
nearly independent of the other modules for low $r$, they may each
converge to sub-parts of different patterns, i.e., the value of $\mu$
would not be identical for the attractors of all the $n_m$ modules.
Thus, the resulting attractor for the entire network corresponds to a 
``chimera" memory state, $\{\xi^{\mu_1} (1), \ldots, \xi^{\mu_{n_m}} (n_m) \}$,
i.e., a spurious pattern comprising fragments of
different stored patterns~\cite{Okane92,Okane93}. 

From the perspective of enhanced robustness of the dynamical attractors 
of the entire network, even more interesting is the behavior of $v_g$
and $v_m$ 
when $r$ is increased further after the modules have become
interconnected appreciably. Over an intermediate range of $p_{min} < p
< p_{max}$, we notice a
non-monotonic variation of both $v_g$ and $v_m$ with respect to $r$.
Fig.~\ref{fig:random_modular_basins} shows that both curves attain a maximum
around  $r_c \sim \frac{n-1}{N-n} \simeq \frac{1}{n_m-1}$, 
where a neuron has the same
number of connections with nodes belonging to its own module as
it has with neurons belonging to different modules.
When the relative number of inter-modular connections are increased 
beyond $r_c$,
the fractional volume of configuration space occupied by the attractors 
corresponding to the stored patterns tend to decrease. 
This implies that the homogeneous network ($r=1$) is actually less robust than
its modular counterpart ($r \simeq r_c$) in terms of global stability 
of the stored attractors.
As $p$ increases beyond $p_{max}$,
both $v_g$ and $v_m$ decrease
at the resulting high loading fraction $p/\langle k \rangle$ through
the generation of a large number of spin-glass states~\cite{Amit89}.
We have explicitly verified that the maximum number of stored patterns
$p_{max}$
beyond which the non-monotonic nature of the variation is lost,
increases when the total number of neurons $N$ is increased, keeping
the overall density of connections, $\langle k \rangle /(N-1)$, and
the number of modules, $n_m$, fixed~\cite{note1}.

\begin{figure}
\begin{center}
\includegraphics[width=0.99\linewidth]{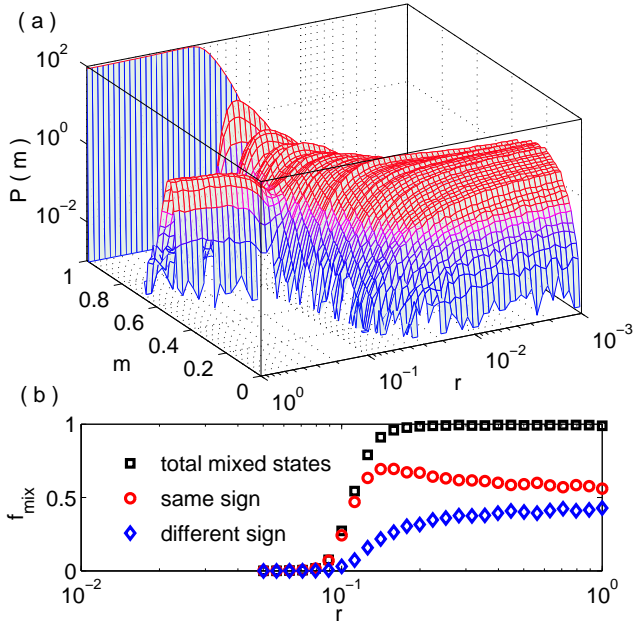}
   \end{center}
 \caption{(a) Distribution of the overlap of the attractors of the network
 dynamics with the stored patterns in a random modular network, at
 different values of the modularity parameter $r$. 
 $P(m)$ is the probability of having overlap $m$.
 Complete overlap
 with the stored patterns ($m=1$) becomes more probable as $r$
 becomes larger than a threshold value. However, at large values of
 $r$, there is a secondary peak around $m_g \sim 0.5$ corresponding to
 mixed states (i.e., linear combination of odd number of stored
 patterns). This peak shows a dip at $r_c \simeq 1/(n_m -1) \sim 0.14$.
 (b) The variation, as a function of $r$, of the fraction of total number 
 of spurious attractors that are mixed states, $f_{mix}$. For $r >
 r_c$, the mixed states account for almost all the attractors not
 corresponding to any of the stored patterns. They can be either
 combinations having
 the same sign (square) or different signs (diamond).
 Results shown for $N=1024$, $n_m = 8$, $\langle k \rangle$ = 120 and
 number of stored patterns, $p = 4$.
 }
 \label{fig:random_modular_overlap}
\end{figure}
For low values of $p$, i.e., $p \leq p_{min}$, both $v_g$ and $v_m$ increase
with $r$ eventually reaching 1 and becoming independent of $r$ once
the connectivity between the modules become appreciable. We find from
our numerical results that
$p_{min} = 2$, independent of the system size $N$ or other model
parameters. This observation helps in identifying the key mechanism
for the non-monotonic variation of $v_g$ with $r$. While at low $r$,
$v_g$
is small because the low connectivity among modules favor the 
chimera states, at very large
$r$ the attractors corresponding to the stored patterns have to
compete with {\em mixed} states. Mixed states are spurious attractors
that correspond
to symmetric combinations of an odd number of stored patterns (e.g.,
$\xi^1 + \xi^2 +\xi^3$) which exist
for all $p > 2$. This is explicitly shown by the distribution of the
overlap, $m$, of the attractors of a network
with any of the $p$ stored patterns
(shown in Fig.~\ref{fig:random_modular_overlap} for $p=4$).
For low values of $r$, the dominance of chimera states result in low
overlap values. 
When the modules become highly inter-connected as 
$r \rightarrow 1$, most randomly chosen initial strings will converge
to a stored pattern resulting in a large
peak at $m=1$ in the overlap distribution. However, we also notice a
smaller peak around $m \simeq 0.5$, which corresponds to 3-pattern mixed
states (which have overlap of 0.5 with each of the three
constituent stored patterns). Note that as $r$ is gradually
decreased from 1, about $r \simeq r_c$ the $m$ distribution shows a sharp dip 
for overlaps around 0.5. This corresponds to an increase in the phase
space volume occupied by the attractors of the stored patterns at the
expense of the mixed states. A similar dip in the distribution is also 
observed for the corresponding overlap around 0.5 for each module
(figure not shown). Thus, the cooperative interactions between the
different modules
not only affect the recall dynamics at the global level, but also
locally within each module.

Fig.~\ref{fig:random_modular_overlap}~(b) shows explicitly
that the attractors not corresponding to any of the stored patterns,
belong almost exclusively to mixed states at high $r$. In
principle, these combinations can be of same sign (e.g.,
$\xi^1+\xi^2+\xi^3$) or different signs (e.g.,
$\xi^1-\xi^2+\xi^3$). The curves corresponding to each of these show
that although the latter has a higher number of possible
combinations, it is the attractors corresponding to the same sign 
combinations which occupy a larger portion of the phase space. This is
a consequence of the Hebbian learning rule, which provides a bias for 
the same sign combinations in preference to the different sign
combinations.

\begin{figure}
\begin{center}
\includegraphics[width=0.99\linewidth]{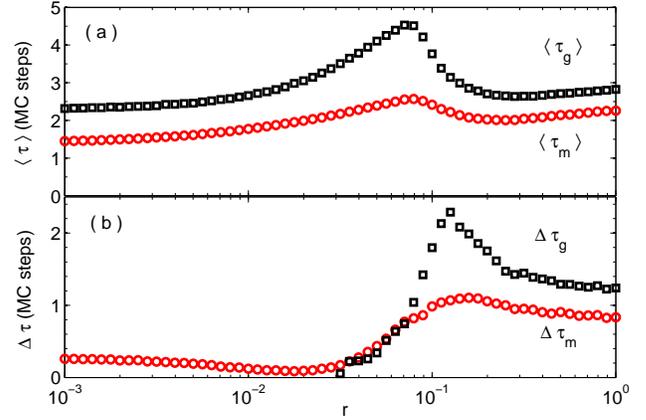}
\end{center}
\caption{(a) The average convergence time $\tau$ to different
attractors in a random modular network, shown for individual modules
($\langle \tau_m \rangle$, circles) and the entire network ($\langle
\tau_g \rangle$, squares). It is measured in
terms of Monte Carlo (MC) steps required to reach a time-invariant state
starting from a random initial configuration. (b) The difference in
the average convergence times (in MC steps) 
to an attractor {\em not} corresponding to
a stored pattern ($m<0.95$) and to one
of the stored patterns, 
$\Delta \tau$. The difference is shown for both
an individual module ($\Delta \tau_m$, circles) and the 
entire network ($\Delta \tau_g$, squares).
The peak close to $r_c \sim 0.14$ corresponds to a significantly faster
convergence to the stored patterns relative to the other attractors.
     Results shown for $N = 1024$, $n_m = 8$, $\langle k \rangle
      = 120$ and $p=4$.} 
\label{fig:convergencetime}
\end{figure}
So far we have discussed the long-time asymptotic properties of the system. 
The
dynamical aspect represented by the time required to reach
equilibrium also exhibits unexpected properties.
Fig.~\ref{fig:convergencetime} shows that the network converges faster
to attractors corresponding to stored patterns as compared to mixed
states (and other attractors that do not have significant overlap with
any of the stored patterns), 
at both the modular and the network level. Moreover, this difference
is slightly enhanced close to $r_c$, the modular configuration where
the basins of the stored patterns cover the largest fraction of the
configuration space. The non-monotonic variation of the convergence time
with decreasing modularity
arises as a result of two competing effects: increasing $r$ decreases
the intra-modular connectivity, resulting in
increasing time for each module to relax to an attractor; on the other
hand, this is accompanied by an increase in the connections between
modules, that eventually causes the entire system to relax faster to
attractors.  
This dynamical picture provides us with a possible clue as to the
enhanced global stability of the attractors corresponding to stored
patterns close to $r_c$. As there is a distinct time-scale separation
between the convergence dynamics at the modular (or local) and at the
global
scale for such networks~\cite{Pan09,Dasgupta09}, 
the state of a specific module may evolve to reach a sub-pattern
corresponding to a part of one of the stored patterns much faster
than the network can converge to an attractor. 
Once this happens, this module biases the convergence of
the other modules connected to it (via Hebbian inter-modular links) to
the pattern to which it has converged. 
This increases the likelihood of convergence of the entire network 
to a particular pattern
through cooperative behavior among the modules, something that is
absent when the modules are disconnected or the network is
homogeneous.

In this paper we have shown that modular organization in the
connection structure of a network of threshold-activated elements 
can result in increased robustness of dynamical attractors associated 
with certain pre-specified
states. These states may represent solutions to
computational tasks or implement memorized patterns of activity. The
modularity of the network allows these states to cover the maximum
volume of its phase space with their basins, an outcome of
cooperative behavior between the convergence or recall dynamics in the
different modules. Our results have special relevance to the question
of how cognitive states arise from interactions between a large number
of brain regions, each comprising many neurons. Neurobiological
evidence exists that cortical activity consists of rapid integration
of signals across brain regions that are in spatially distinct
locations and which occurs in a self-organized manner through
interactions between the elements of the network of brain 
areas~\cite{Sporns06}. The empirical observation of modular cortical
organization and the occurrence of distinct, persistent activity
patterns corresponding to attractor dynamics raises the intriguing
possibility that evolution may have selected modularity because of the
robustness it imparts to the underlying system. Future extensions of
the work reported here may involve considering the effect of noise,
i.e., investigating the recall dynamics at a finite temperature.
Another possibility is to investigate the role of hierarchical arrangement 
of modules that have recently been reported in different biological
systems~\cite{Ravasz02,Pan08}, including the
brain~\cite{Ferrarini08,Meunier09b}.
Our results may also potentially be used to understand why attractor
networks with small-world connection topology show a small increase in
global stability relative to random networks, although the local
stability of stored patterns are
unaffected~\cite{Morelli04,McGraw03,Kim04,Oshima07}.
%

\vspace{0.5cm}
We would like to thank R.~K. Pan for helpful discussions. This work is
supported in part by CSIR, UGC-UPE, IMSc Associates program and
IMSc Complex Systems (XI Plan) Project.


\end{document}